\def\aa{{A\&A}}

\def\aj{{AJ}}
\def\annrev{{ARA\&A}}
\def\apj{{ApJ}}

\def\prd{{Phys. Rev. D}}
\def\prl{{Phys. Rev. Lett.}}

\def\edth{\;\raise1.0pt\hbox{$'$}\hskip-6pt\partial\;}
\def\baredth{\;\overline{\raise1.0pt\hbox{$'$}\hskip-6pt
\partial}\;}
\def\bi#1{\hbox{\boldmath{$#1$}}}

\documentclass[cup5b]{caps}
\usepackage{epsf}
\usepackage{graphicx}
\usepackage{amssymb}
\usepackage{ociwsymp2}
\HeadText{M. Zaldarriaga}

\begin{document}

\pagenumbering{arabic}

\author[]{MATIAS ZALDARRIAGA\\ Astronomy \& Physics Departments, 
Harvard University}

\chapter{The Polarization of the Cosmic Microwave Background}

\begin{abstract}

We summarize the physical mechanism by which the
cosmic microwave background acquires a small degree of polarization. We
 discuss the imprint left by gravitational waves and the use of
polarization as a test of the inflationary paradigm. We  discuss some
physical processes that affect the cosmic microwave background polarization 
after recombination, such as gravitational lensing and the 
reionization of the Universe.

\end{abstract}

\section{Introduction}

Since its discovery in 1965 the cosmic microwave background (CMB) has
been one of the pillars of the Big Bang model (Penzias \& Wilson 1965).
The various measurements of its spectrum, in particular by the {\it COBE}/FIRAS
instruments firmly established the hot big bang model as the basis of
our understanding of cosmology (Mather et al. 1994).

The study of the spectrum was followed by the detection of tiny
anisotropies in the CMB temperature, first by {\it COBE}/DMR (Smoot
et al. 1992), and then
by a variety of more sensitive experiments with better angular
resolution\footnote{A compilation of up to date results can be found
at http://www.hep.upenn.edu/\~{}max/ . We apologize for 
not giving an up to date list  of 
all experiments but that made us exceed the page limit.}. 
The anisotropies, a natural consequence
of the structure formation process, contain a wealth of information
about the cosmological model. They depend on the matter content of the
Universe and on the physical process that created the tiny seeds that
grew under gravity to form the structure in the Universe around us.
Moreover, the structure formation process leaves its imprint on
the CMB through secondary effects such as gravitational lensing, the
Sunyaev-Zel'dovich effect (Sunyaev \& Zel'dovich 1972), or the
Ostriker-Vishniak effect (Ostriker \& Vishniak 1986).  

The detailed study of the temperature anisotropies has taken
the field into the era of ``precision cosmology.'' Ever more sensitive
temperature experiments have so far confirmed our theoretical 
picture. These studies have revealed the presence of acoustic
peaks in the power spectrum and of a damping tail on small scales. The
comparison of  measurements and theory, most recently the results of {\it WMAP}
(Hinshaw et al. 2003),  have led to very narrow constraints on several of 
the cosmological parameters. Experiments already under way or being 
constructed, such as the {\it Planck} satellite, will tighten constraints even 
further.  We warn the reader that this article was written prior to the 
release of the {\it WMAP} results; references to them were added just before 
publication.

\begin{figure*}[t]
\includegraphics[width=1.00\columnwidth]{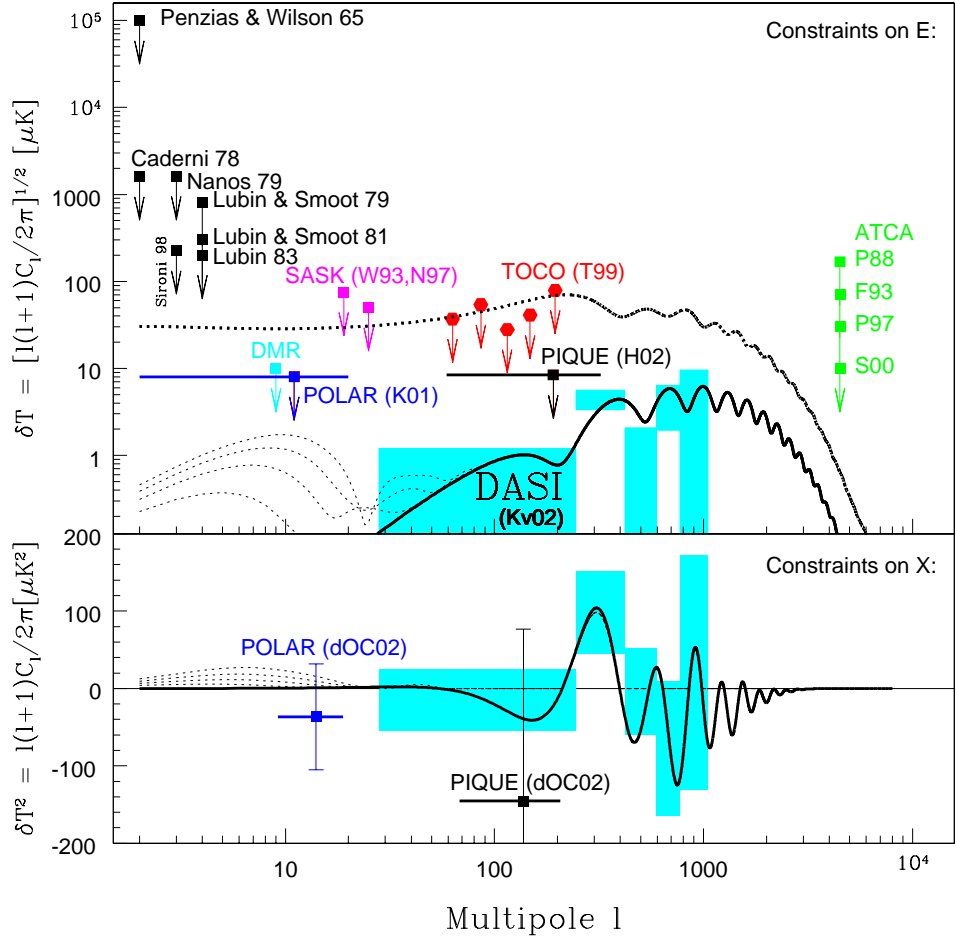}
\vskip 0pt \caption{
Summary of polarization measurements and upper limits before the 
release of {\it WMAP} results, kindly provided by Angelica de~Oliveira-Costa 
(de~Oliveira-Costa et al 2002b). The top (bottom) panel shows results for the 
$E$ (cross-correlation) power spectra. 
\label{exper}}
\end{figure*}

The next big goal for CMB experimentalists was the detection of the
even smaller CMB polarization anisotropies. This was accomplished for
the first time by the beautiful DASI experiments 
(Kovac et al. 2002; Leitch at al.
2002). Since the first detection of the CMB, there have been many
theoretical studies of the expected polarization properties (e.g., Rees
1968; Polnarev 1985) and numerous attempts to
measure or put upper limits on it.
Figure \ref{exper} provides a summary of experimental
results up to the time of writing. 
``Modern'' experiments include POLAR (Keating et al. 2001) 
and PIQUE (Hedman et al. 2002), which set the most stringent upper
limits before the DASI detection, 
roughly around 10 $\mu \rm K$. The correlation between temperature
and polarization was searched for by comparing PIQUE with
Saskatoon (de~Oliveira-Costa et al. 2003a,b), was detected by DASI, and has 
now been measured with exquisite signal-to-noise ratio over a wide range of 
scales by {\it WMAP} (Kogut et al. 2003). 

DASI opened 
a new window into the early Universe. Polarization is sensitive
to most of the parameters in the cosmological model. The way it
depends on them is often different than the way temperature
anisotropies do. As a result, accurate measurements of the CMB
polarization will improve the determination of many cosmological parameters 
(Zaldarriaga, Spergel, \& Seljak 1997; Eisenstein, Hu, \& Tegmark 1999). 
Moreover, polarization will provide an excellent test bed for
consistency check on parts of the model that at present we take
for granted.

The big push toward polarization, however, comes from its potential as
a detector of a stochastic background of gravitational waves (GW). It has
been shown that the pattern of polarization ``vectors'' on the sky
will be different --- it will have a curl-like component --- if there is a
stochastic background of GW (Kamionkowski, Kosowsky \&
Stebbins 1997a; Seljak \& Zaldarriaga 1997) . Inflation models 
predict the existence of a stochastic background of 
GW. A detection would provide the ``smoking gun'' for 
inflation. The amplitude of the GW in this background is directly 
proportional to the square of the energy scale of inflation, so that a 
detection of polarization would nail down key parameters of the inflationary 
model (e.g., Dodelson, Kinney, \& Kolb 1997; Kinney 1998). 

In this article we will review the mechanism by which polarization is
generated and discuss what can be learned from measurements
of polarization. In \S~\ref{origin} we will briefly review
the physics of temperature anisotropies, although the reader is
encouraged to go to one of the many reviews in the literature for
further details. In \S~\ref{polarization} we will discuss 
how polarization gets generated during recombination and to what it is 
sensitive. 
In \S~\ref{gw} we will discuss the imprint of
GW. In \S~\ref{afterrec} we will review some of the
physical processes that can affect the polarization signal
after recombination. We conclude in \S~\ref{conc}.

\section{Temperature Anisotropies}\label{origin}

In this section we will summarize the relevant facts about hydrogen
recombination and temperature anisotropies. It is not our intention to
provide an extensive review of these topics.  The interested reader should
consult other reviews (e.g., Hu \& Dodelson 2002; Hu 2003). A discussion
on the relevance of CMB studies for particle physics can be found in
Kamionkowski \& Kosowsky (1999).

\subsection{Hydrogen Recombination and Thomson Scattering}\label{Thomsom}

The most abundant element in our Universe is hydrogen, and its
ionization state has profound consequences on the CMB. The temperature
of the CMB increased linearly with redshift, $T\propto (1+z)$.  The
interaction between the CMB photons and the hydrogen atoms kept
hydrogen ionized until a redshift of $z\approx 1000$. 
At this time, corresponding to a conformal time of $\tau_R\approx 110\ 
(\Omega_m h^2)^{-1/2}\ {\rm Mpc}$, there are not enough
energetic photons in the CMB to keep hydrogen ionized, 
so it recombines. Conformal time
is defined by $a\ d\tau= dt$, with $a$ the expansion factor and $t$
the physical time. It is useful because a null geodesic is
simply given by $d\tau=dx$, where $x$ are comoving coordinates ($a\
dx=dr$, where $r$ is the physical distance).
For example the fact that $\tau_R\approx 110\ 
(\Omega_m h^2)^{-1/2}\ {\rm Mpc}$ means a photon traveling on a
straight line since the Big Bang until recombination would have traveled
between two points that {\it today} are separated by a physical
distance of $110\ (\Omega_m h^2)^{-1/2}\ {\rm Mpc}$. 

\begin{figure*}[t]
\includegraphics[width=1.00\columnwidth]{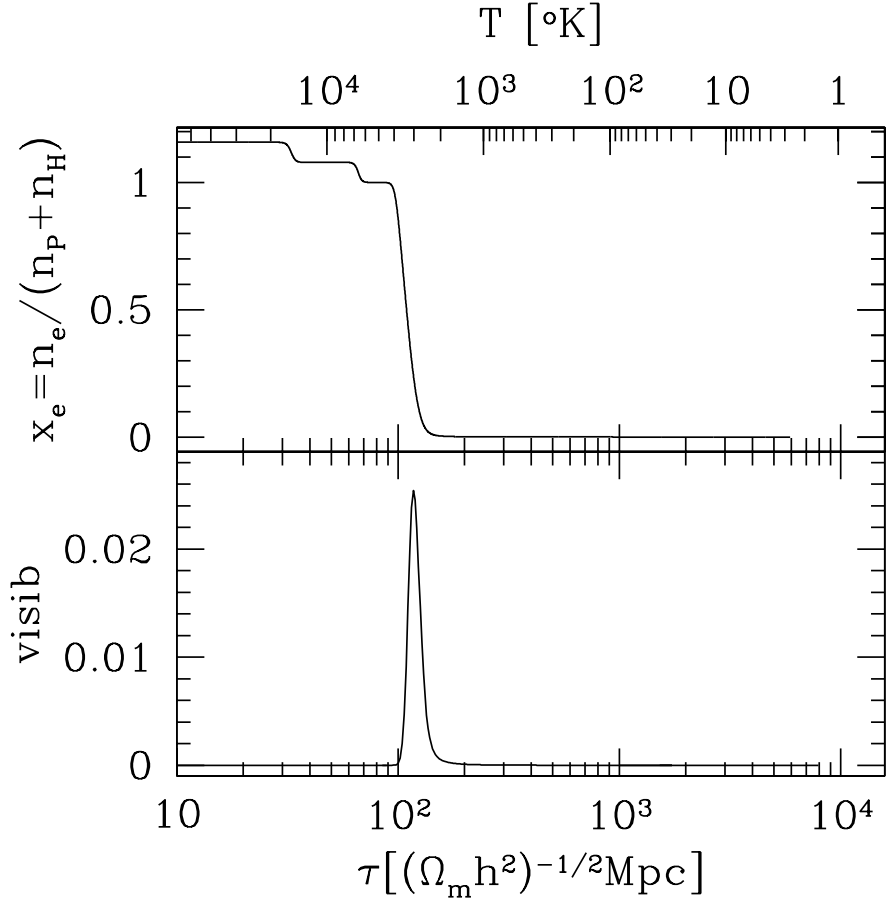}
\vskip 0pt \caption{
In the top panel we show the ratio of number densities of free
electrons and baryons. At early times the fraction is bigger than 
one because it includes electrons coming from helium. The bottom
panel shows the visibility function that gives
the probability a photon we observe last scattered at a particular
position along the line of sight. The x-axis is labeled using
conformal coordinates on the bottom and the temperature of the CMB on
the top. 
\label{fighydrec}}
\end{figure*}

The fraction of free electrons coming from hydrogen and helium
is shown in Figure \ref{fighydrec} (it was calculated using 
RECFAST of Seager, Sasselov, \& Scott 1999). 
One of the main points to take away from
the figure is that recombination happens rather fast. The width of
recombination is $\delta \tau_{R} / \tau_{R} \approx 0.1-0.2$.

The other process we need to consider to understand how anisotropies
are generated is Thomson scattering. When
hydrogen is ionized the CMB photons can scatter with  free
electrons. This process conserves the number of photons and does not change
their energy (because we are working in the limit of infinite
electron mass). The mean free path for Thomson scattering in comoving
$\rm Mpc$ is,
$\lambda_T=(a n_e \sigma_T)^{-1}$
$= 4 \ {\rm Mpc}\ {1\over x_e} ({0.0125\over \Omega_b h^2})
  ({0.9 \over 1-Y_p/2})({1000\over (1+z)})^2$, where
$n_e$, $x_e$, and $Y_p$ are the electron density, fraction  of free electrons, 
and the primordial helium mass fraction, respectively.

Before recombination 
($x_e\approx 1$) the mean free path was much
smaller than the horizon at the time [$\tau_R\approx 110 \ (\Omega_m
h^{2})^{-1/2}\, {\rm
Mpc}$]. Photons were hardly able to travel without scattering, a
regime called ``tight coupling.'' After recombination 
($x_e\approx 3\times 10^{-4}$) the mean free path becomes much
larger than the horizon, so photons can travel in a straight line
to our detectors. The Universe has become transparent for CMB
photons, a process also called decoupling. 


In the bottom panel of 
Figure \ref{fighydrec} we show the visibility function, which gives
the probability a photon we observe last scattered at a particular
position along the line of sight.  
The function is strongly peaked. After recombination
the photons no longer scatter, so the visibility function is almost
zero. Before recombination the scatterings were so frequent that for
most photons the last scattering occurs right around
recombination. Thus, the CMB photons 
are coming from a very narrow range in distance, given by the
width of the last-scattering surface, $\delta\tau_R/\tau_R \approx
0.1$. The collection
of these regions in all directions on the sky form a very thin sphere
of radius $D_{LSS}=\tau_0-\tau_R\approx 6000 (\Omega h^2)^{-1/2}{\rm Mpc}$,
called the last-scattering surface (LSS). 

\subsection{Review of Temperature Anisotropies}

Now we are in a position to understand the origin of the CMB
anisotropies. There are several reasons why the temperature we observe
in different directions is not exactly the same. Differences in the density
of photons across the last-scattering surface will lead
to differences in the observed intensity as a function of position on the sky. 
If the density of photons is higher in a particular
region, we see this as a hot spot.  These
same density differences across the Universe lead to gravitational
potential differences. If photons climb (go down) a potential well to get to us,
they get redshifted (blueshifted), 
and this decreases (increases) the observed temperature. The
gravitational potential differences create forces that cause motions, 
and these motions create shifts in frequency due to the Doppler effect.
Thus, if photons come to us from a region that is moving toward (away from) us, 
they get blueshifted (redshifted), which we observe as a higher (lower)
temperature.
Finally, if the gravitational potentials are changing with time, the
energy of photons changes along the way, leading to the so
called integrated Sachs-Wolfe (ISW) effect (Sachs \& Wolfe 1967). For a detailed discussion
of all the relevant equations in both synchronous and conformal gauges, 
see Ma \& Bertschinger (1995), and for more details as to how to treat
polarization and GW, see Hu et al. (1998). 

In the Newtonian gauge, the temperature fluctuations $\Delta_T=\delta
T/T$ observed in direction $\bi {\hat n}$ can be written as
\begin{eqnarray}\label{integsol}
  (\Delta_T+\psi)(\hat{\bi n})=({\delta_{\gamma} \over 4}+\psi)|_{\tau_R}
+ \hat {\bi n} \cdot {\bi v}_b|_{\tau_R}+ \int d\tau^{\prime}
  e^{-\kappa(\tau_0,\tau^{\prime})}
  (\dot{\phi}+\dot{\psi}).
\end{eqnarray}
We have introduced $\delta_\gamma=\delta\rho_\gamma/\rho_\gamma$,  
the fractional energy density fluctuations in the CMB, 
${\bi v}_b$ the baryon velocity, $\phi$ and
$\psi$ the two gravitational potentials in terms of which the metric
fluctuations are given by $ds^2=a^2(\tau)[-(1+2\psi) d\tau^2+(1-2\phi) d{\bi
x}^2]$, and the optical depth
$\kappa(\tau_2,\tau_1)=\int_{\tau_2}^{\tau_1} a n_e x_e d\tau$. The
first term is usually referred to as the monopole contribution, while
the second is the Doppler or dipole contribution. 

In Equation \ref{integsol} we
have not included the terms coming from polarization 
because they are sub-dominant. We have 
approximated the visibility function by a $\delta$-function; that is
why the first two terms are evaluated at recombination. 
The different terms describe the physical
effects mentioned above. From left to right, they correspond to photon 
energy density fluctuations, gravitational potential redshifts, Doppler
shifts, and the ISW effect. Note that the ISW effect is an integral along the
line of sight not constrained to recombination. 

To be able to calculate the power spectrum of the anisotropies, we need
to evaluate the different terms in Equation \ref{integsol}. Because
we are dealing with small perturbations and linear theory, 
perturbations are usually expanded in Fourier modes, and the 
anisotropies are calculated for each
Fourier mode individually. The contributions of individual modes 
are then added to calculate the power spectrum.  

As discussed in the previous section, before recombination Thomson
scattering was very efficient. As a result it is a good approximation
to treat photons and baryons as a single fluid. 
This treatment is called the tight-coupling approximation and will allow
us to evolve the perturbations until recombination to calculate the
different terms in Equation \ref{integsol}.

The equation for the photon density perturbations for one Fourier
 mode of wavenumber $k$ is that of a forced and damped
harmonic oscillator (e.g., Seljak 1994; Hu \& Sugiyama 1995):
\begin{eqnarray}\label{fdosc}
  \ddot{\delta_\gamma}+{\dot{R} \over (1+R)} \dot{\delta_{\gamma}}
  +k^2 c^2_{s} \delta_{\gamma}&=&F \nonumber \\
F&=&4[\ddot \phi +{\dot R \over (1+R)} \dot \phi - {1 \over 3} k^2
  \psi] \nonumber \\
\dot \delta_\gamma&=& -{4 \over 3 } k v_\gamma + 4 \dot \phi.
\end{eqnarray}
The photon-baryon fluid can sustain acoustic oscillations.
The inertia is provided by the baryons, while the pressure is provided
by the photons. The sound speed is $c_s^2=1/3(1+R)$, with
$R=3\rho_b/4\rho_{\gamma}=31.5\ (\Omega_b h^2)
(T/2.7)^{-4}[(1+z)/10^3]^{-1}$.  As the baryon
fraction goes down, the sound speed approaches $c_s^2\rightarrow 1/3$. 
The third equation above is the continuity equation. 

As a toy problem, we will solve Equation \ref{fdosc} under 
some simplifying assumptions. If we
consider a matter-dominated universe, the driving force becomes a
constant, $F=-4/3 k^2 \psi$, 
because the gravitational potential remains constant in time. 
We neglect anisotropic stresses so that $\psi=\phi$, and,  
furthermore, we neglect the time dependence of $R$.  
Equation \ref{fdosc} becomes
that of a harmonic oscillator that can be trivially solved. This is
a very simplified picture, but it captures most
of the relevant physics we want to discuss. More elaborate
approximation schemes can be found in the literature. They allow the
calculation of the power spectrum with an accuracy of roughly 10\%
(Seljak 1994; Hu \& Sugiyama 1995; Weinberg 2001a,b; Mukhanov 2003).

To obtain the final solution we need to specify the initial
conditions. We will restrict ourselves to adiabatic initial
conditions, the most natural outcome of inflation. In our context this
means that initially $\phi=\psi=\phi_0$, $\delta_\gamma=-8/3 \phi_0$, 
and $v_\gamma=0$. We have denoted $\phi_0$ the initial amplitude of
the potential fluctuations. We will take $\phi_0$ to be a Gaussian 
random variable with power spectrum $P_{\phi_0}$. 

We have made enough approximations that the evaluation of the sources
in the integral solution has become trivial. The solution for
the density and velocity of the photon fluid at recombination are:
\begin{eqnarray}\label{solution1}
({\delta_{\gamma} \over 4}
+\psi) |_{\tau_R}&=&{\phi_0 \over 3}(1+3R)\cos (k c_s \tau_R)-\phi_0 
R \nonumber \\
v_\gamma |_{\tau_R}&=& -\phi_0 (1+3R) c_s \sin(k c_s \tau_R).
\end{eqnarray}
Equation \ref{solution1} is the solution for a single Fourier
mode. All quantities have an additional spatial dependence
($e^{i\bi k \cdot \bi x}$), which we have not included to make the
notation more compact. 
With that additional term the solution of Equation \ref{integsol} becomes
\begin{eqnarray}
  \Delta_{T}(\hat {\bi n})&=&e^{ikD_{LSS}\cos\theta} S \nonumber \\
S&=&\phi_0 {(1+3R) \over 3}
[\cos (k c_s \tau_R)- {3 R \over (1+3R)} \nonumber \\
&& - i \sqrt{3 \over 1+R} 
\cos \theta \sin(k c_s
  \tau_R)],
\label{integapprox1mode}
\end{eqnarray}
where we have neglected the $\psi$ on the left-hand side because it is
a constant independent of $\bi{\hat n}$. We have also
ignored the ISW contribution.  
We have introduced $\cos \theta$, the cosine of the angle between
the direction of observation and the wavevector $\bi k$; for example, 
$\bi k \cdot \bi x=k D_{LSS} \cos \theta$ . The term
proportional to $\cos \theta$ is the Doppler contribution. 

Once the temperature perturbation produced by one Fourier mode has
been calculated, we need to expand it into spherical harmonics to
calculate the $a_{lm}= $ $\int d\Omega $ $ Y_{lm}^*(\bi {\hat n})$ $
\Delta T(\bi {\hat n})$. The power spectrum of temperature
anisotropies is expressed in terms of the $a_{lm}$ coefficients as
$C_{Tl}=\sum_m |a_{lm}|^2$. The contribution to $C_{Tl}$ from each
Fourier mode is weighted by the amplitude of primordial fluctuations
in this mode, characterized by the power spectrum of $\phi_0$, 
$P_{\phi_0}=A k^{n-4}$. We will take the power-law index to be $n=1$
in our approximate formulas. In practice, fluctuations on angular scale
$l$ receive most of their contributions from wavevectors around
$k^*=l/D_{LSS}$, so roughly the amplitude of the power spectrum 
at multipole $l$ is given by the value of the sources in Equation
\ref{solution1} at $k^*$.

After summing the contributions from all modes, the power spectrum is
roughly given by 
\begin{eqnarray}\label{approxcl}
  l(l+1)C_{Tl}&\approx&A \{[{(1+3R)\over 3} \cos(k^*c_s\tau_R)-R]^2+ 
\nonumber \\
&& {(1+3R)^2
  \over 3} c_s^2  
  \sin^2(k^*c_s\tau_R)\}.
\end{eqnarray}

\begin{figure*}[t]
\includegraphics[width=1.00\columnwidth]{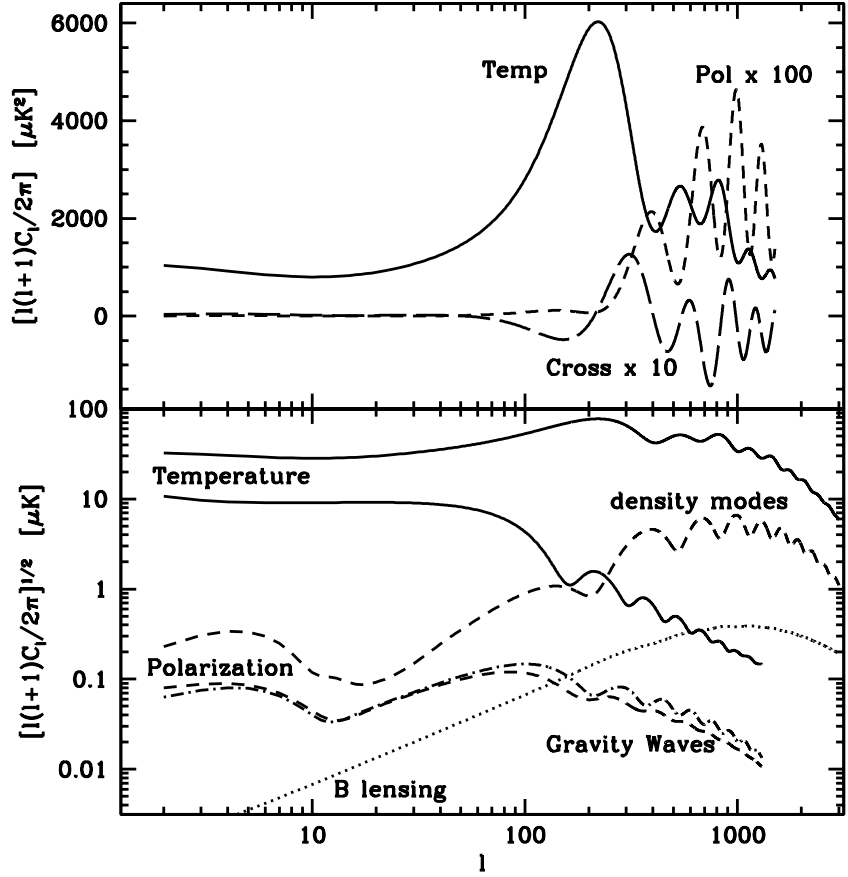}
\vskip 0pt \caption{
     Temperature, polarization, and cross-correlation
     power spectrum produced by density perturbations in the fiducial
     $\Lambda$CDM model ($\Omega_m=0.3$, $\Omega_\Lambda=0.7$, $h=0.7$,
     $\Omega_bh^2=0.02$). For display purposes the polarization
     (cross-correlation) power
     spectrum was multiplied by a factor of 100 (10).
    \label{figclcdm}}
 \end{figure*}

Equation \ref{approxcl} can be used to understand the basic
features in the CMB power spectra shown in Figure \ref{figclcdm}. 
The baryon drag on the photon-baryon fluid reduces its sound speed below 
$1/3$ and makes the monopole
contribution dominant [the one proportional to $\cos(k^*c_s\tau_R$]. 
Thus, the $C_{Tl}$ spectrum peaks where the
monopole term peaks, $k^*c_s\tau_R=\pi,2\pi,3\pi,\cdots$, which
correspond to $l_{peak}=n\pi D_{LSS}/c_S\tau_R$. More detailed
discussions of the physics of the acoustic peaks can be found in
reviews such as that by Hu \& Dodelson (2002) or Hu (2003).

It is very important to understand the origin of the acoustic
peaks. In this model the Universe is
filled with standing waves; all modes of wavenumber $k$ are in
phase, which leads to the oscillatory terms. The sine and cosine in Equation
\ref{approxcl} originate in the time dependence of the modes. 
Each mode $l$
receives contributions preferentially from Fourier modes of a
particular wavelength $k^*$ (but pointing in all directions), 
so to obtain peaks in $C_l$,  
it is crucial that all modes of a given $k$ are in phase.
If this is not the case, the features in the $C_{Tl}$ spectra will be
blurred and can even disappear. This is what happens when one considers the
spectra produced by topological defects (e.g., Pen, Seljak, \& Turok 1997). 
The phase coherence of all
modes of a given wavenumber 
can be traced to the fact that perturbations were produced very
early on and had wavelengths larger than the horizon during many expansion
times. 

There are additional physical effects we have neglected. 
 The Universe was radiation dominated early on,
and modes of wavelength smaller and bigger than the horizon at
matter-radiation equality behave differently. During the
radiation era the perturbations in the photon-baryon
fluid are the main source for the gravitational potentials which decay once a mode enters into the horizon. The
gravitational potential decay acts as a driving force for the oscillator
in Equation \ref{fdosc}, so a feedback loop is established. As a
result, the acoustic oscillations for 
modes that entered the horizon before matter-radiation
equality have a higher amplitude. In the $C_{Tl}$ spectrum the
separation between modes that experience this feedback and those
that do not occur at $l\sim D_{LSS}/ \tau_{eq}$, where $\tau_{eq}\approx 16
(\Omega_m h^2)^{-1} \, {\rm Mpc}$ is the
conformal time of matter-radiation equality. Larger $l$ values
receive their contributions from modes that entered the horizon
before matter-radiation equality. Finally, when a mode is inside the
horizon during the radiation era the gravitational potentials decay. 

There is a competing effect, Silk damping (Silk 1968),  that reduces the 
amplitude of the large-$l$ modes. The photon-baryon fluid
is not a perfect fluid. Photons have a finite mean free path and
thus can random walk away from the peaks and valleys 
of the standing waves.
Thus, perturbations of wavelength comparable to or smaller than the
distance the photons can random walk get damped. This
effect can be modeled by multiplying Equation \ref{integapprox1mode}
by $\exp(-k^2/k_s^2)$, with $k^{-1}_{s}\propto \tau_R^{1/2} (\Omega_b
h^2)^{-1/2}$. Silk damping is important for multipoles of order $l_{Silk}\sim
k_s D_{LSS}$.

Finally, the last-scattering surface has a finite width (Fig.
\ref{fighydrec}). 
Perturbations
with wavelength comparable to this width get smeared out due to
cancellations along the line of sight. This
effect introduces an additional damping with a characteristic scale
$k^{-1}_w \propto \delta\tau_R$.

\section{Polarization}\label{polarization}

In this section we will describe how polarization is characterized and
how it is generated by density perturbations. We will stress the
similarities and differences with temperature anisotropies.  We will
focus on polarization generated by density perturbations and leave the
signal of GW for a later section. The physics of
polarization has been previously reviewed in Hu \& White (1997),
Kosowsky (1999), and Hu (2003).

\subsection{Characterizing the Radiation Field}

The aim of this part is to summarize the mathematical tools needed to
describe the CMB anisotropies. 
The anisotropy field is characterized by a $2\times 2$ intensity 
tensor $I_{ij}$. For convenience, we normalize this tensor so it represents 
the fluctuations in
units of the mean intensity ($I_{ij}=\delta I /I_0$). 
The intensity tensor is a
function of direction on the sky, $\hat{\bi{n}}$, and  two directions
perpendicular to $\hat{\bi{n}}$ that are  used to define its components
(${\bf \hat e}_1$,${\bf \hat e}_2$).
The Stokes parameters $Q$ and $U$ are defined as
$Q=(I_{11}-I_{22})/4$ and $U=I_{12}/2$, while the temperature 
anisotropy is
given by $T=(I_{11}+I_{22})/4$ (the factor of $4$ relates fluctuations
in the intensity with those in the temperature, $I\propto T^4$).  
When representing polarization using ``rods'' in a map, 
the magnitude is given by $P=\sqrt{Q^2+U^2}$, and the 
orientation makes 
an angle $\alpha={1\over 2}\arctan({U/Q})$ with ${\bf \hat e}_1$.
In principle the fourth  
Stokes parameter $V$ that describes circular polarization is needed, 
but we ignore it because it cannot
be generated through Thomson scattering, so the CMB is not expected to
be circularly polarized.   
While the temperature is invariant
under a right-handed rotation in the plane perpendicular to direction
$\hat{\bi{n}}$,
$Q$ and $U$ transform under rotation by an angle $\psi$ as
\begin{equation}
(Q\pm iU)'(\hat{\bi{n}})=e^{\mp 2i\psi}(Q\pm iU)(\hat{\bi{n}}),
\end{equation}
where ${\bf \hat e}_1^{\prime}=\cos \psi\ {\bf \hat e}_1+\sin\psi\ 
{\bf \hat e}_2$ 
and ${\bf \hat e}_2^{\prime}=-\sin \psi\ {\bf \hat e}_1+\cos\psi\ 
{\bf \hat e}_2$. The quantities $Q\pm iU$ are said to be spin 2. 

We already mentioned  that the statistical properties of the
radiation field are usually described in terms the spherical harmonic
decomposition of the maps. This basis, basically the Fourier basis, is
very natural because the statistical properties of 
anisotropies are rotationally invariant.  
The standard spherical harmonics are not the
appropriate basis for $Q\pm iU$ because they are spin-2 variables, but 
generalizations (called $_{\pm 2} Y_{lm}$) exist. We can expand 
\begin{eqnarray}
(Q\pm iU)(\hat{\bi{n}})&=&\sum_{lm} 
a_{\pm 2,lm}\;_{\pm 2}Y_{lm}(\hat{\bi{n}}). 
\label{Pexpansion}
\end{eqnarray}
$Q$ and $U$ are defined at each direction $\hat {\bi{n}}$
with respect to the spherical coordinate system $(\hat{{\bf e}}_\theta,
\hat{{\bf e}}_\phi)$ (Kamionkowski, Kosowsky, \&
Stebbins, 1997b; Zaldarriaga \& Seljak 1997).
To ensure that $Q$ and $U$ are real, 
the expansion coefficients 
must satisfy $a_{-2,lm}^*=a_{2,l-m}$. The equivalent relation 
for the temperature coefficients is $a_{T,lm}^*=a_{T,l-m}$.

Instead of $a_{\pm 2,lm}$, it is convenient to introduce their
linear combinations 
$a_{E,lm}=-(a_{2,lm}+a_{-2,lm})/2$ and
$a_{B,lm}=i(a_{2,lm}-a_{-2,lm})/2$ (Newman \& Penrose 1966). 
We define two quantities in real space,
$E(\hat{\bi{n}})=\sum_{l,m}a_{E,lm} \ Y_{lm}(\hat{\bi{n}})$ and 
$B(\hat{\bi{n}})=\sum_{l,m}a_{B,lm} \ Y_{lm}(\hat{\bi{n}})$. $E$ and
$B$ completely specify the linear polarization field. 

The temperature is a
scalar quantity
under a rotation of the coordinate system,
$T^{\prime}(\hat{\bi{n}}^{\prime}={\bf \cal R} \hat{\bi{n}})
=T(\hat{\bi{n}})$, where  $\bf {\cal
R}$ is the rotation matrix. We denote with a
prime the quantities in the transformed coordinate system. While $Q\pm
i U$ are spin 2,
$E(\hat{\bi{n}})$ and $B(\hat{\bi{n}})$ are 
invariant under rotations. Under parity, however, $E$ and $B$ behave
differently, $E$ remains unchanged, while $B$ changes sign. 

\begin{figure*}[t]
\includegraphics[width=1.00\columnwidth]{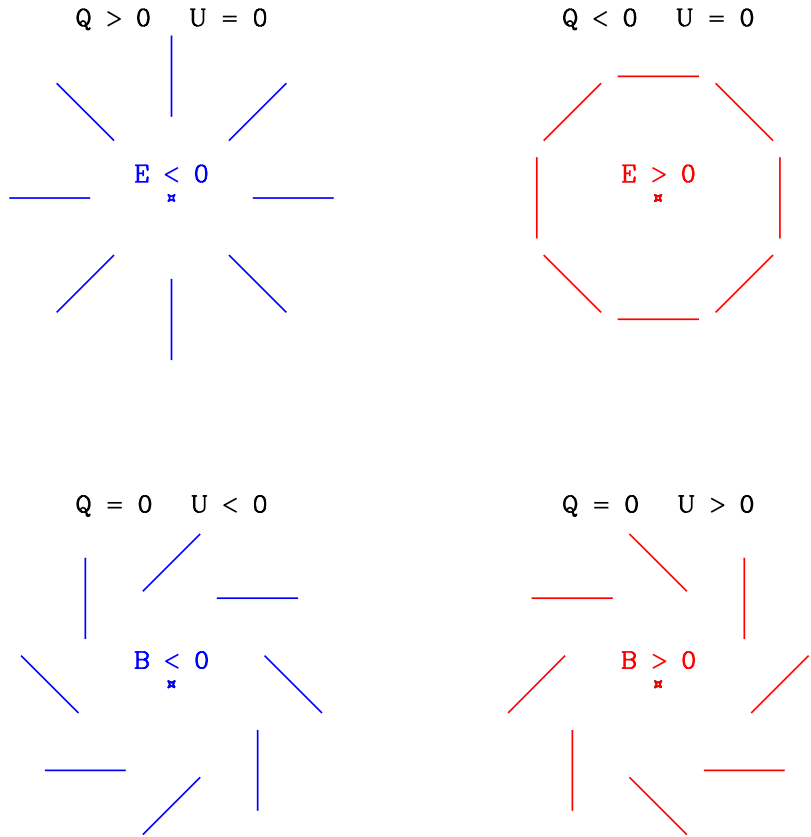}
\vskip 0pt \caption{
     Example of $E$-type and $B$-type 
       patterns of polarization. Note that if reflected across 
       a line going through the center the $E$ patterns are 
       unchanged, while the $B$ patterns switch with one another.
\label{eb1}}
\end{figure*}

As an illustration, in Figure \ref{eb1} we
show polarization patterns that have positive and negative $E$ and
$B$. It is clear that $B$ patterns are ``curl-like,'' having different
properties than the $E$ patterns under parity transformation. When
reflected across a line going through the center, the $E$ patterns
remain unchanged, whereas the $B$ patterns change from positive to
negative. We want to
stress at this point that whether a polarization field has an $E$ or
$B$ component is a property of the pattern of polarization rods {\it around}
a particular point, and not at the point itself. 
In this sense $E$ and $B$ are not local
quantities.

There is a clear analogy between vector fields and
polarization fields with regard to their geometrical properties.
The polarization field is a spin-2 field, which means that
if one rotates the coordinate system by $180^{\circ}$ one goes back to the same
components of the field, as opposed to $360^{\circ}$ needed for vector
fields. However, conceptually things are very similar, with $E$ and $B$
playing the roles of the gradient and curl parts. See Bunn et al. (2003)
for a more detailed description of similarities and
differences between vectors and polarization fields.

To characterize the statistics of the CMB perturbations, 
only four power spectra are needed, 
those for $T$, $E$, $B$ and the cross correlation between $T$ and $E$.
The cross correlation between $B$ and $E$ or $B$ and 
$T$ vanishes if there are no parity-violating interactions because 
$B$ has the opposite parity to $T$ or $E$. Examples of models where
this is not true were presented in Lue, Wang, \& Kamionkowski (1999). 
The power spectra are defined as the rotationally invariant quantities
$C_{Tl}={1\over 2l+1}\sum_m \langle a_{T,lm}^{*} a_{T,lm}\rangle$, 
$C_{El}={1\over 2l+1}\sum_m \langle a_{E,lm}^{*} a_{E,lm}\rangle$, 
$C_{Bl}={1\over 2l+1}\sum_m \langle a_{B,lm}^{*} a_{B,lm}\rangle$, 
and
$C_{Cl}={1\over 2l+1}\sum_m \langle a_{T,lm}^{*}a_{E,lm}\rangle$.
The brackets $\langle \cdots \rangle$ denote ensemble averages.

We will not discuss how to define correlation functions in real space
for the polarization field. The reader is referred to the literature
(e.g., Kamionkowski et al. 1997b; Zaldarriaga 1998; Tegmark \& de~Oliveira-Costa 2001). We only want to point out that the spin nature
of the Stokes parameters needs to be considered to properly define
correlation functions. 

\subsection{The Physics of Polarization}

Polarization is generated by Thomson scattering between photons and
electrons, which means that polarization cannot be generated after
recombination (except for reionization, which we will discuss later). 
But Thomson scattering is
not enough. The radiation incident on the electrons must also be
anisotropic. In fact, its intensity needs to have a quadrupole moment. This
requirement of having both Thomson scattering and anisotropies is what
makes polarization relatively small. After recombination, anisotropies
grow by free streaming, but there are no scattering to generate
polarization. Before recombination there were so many scatterings
that they erased any anisotropy present in the photon-baryon fluid. 

In the context of anisotropies induced by density perturbations,
velocity gradients in the photon-baryon fluid are responsible for the
quadrupole that generates polarization.  Let us consider a scattering
occurring at position $\bi x_0$: the scattered photons came from a
distance of order the mean free path ($\lambda_T$) away from this
point. If we are considering photons traveling in direction $\bi
{\hat n}$, they roughly come from $\bi x=\bi x_0 + \lambda_T \bi {\hat
  n}$. The photon-baryon fluid at that point was moving at velocity
$\bi v(\bi x)\approx \bi v(\bi x_0)+\lambda_T {\bi {\hat n}}_i
\partial_i {\bi v}(\bi x_0)$. Due to the Doppler effect the
temperature seen by the scatterer at $\bi x_0$ is $\delta T(\bi
x_0,\bi {\hat n})= \bi {\hat n}\cdot [\bi {v}(\bi x)-\bi v(\bi
x_0)]\approx\lambda_T {\bi {\hat n}}_i{\bi {\hat n}}_j \partial_i {\bi
  v}_j(\bi x_0)$, which is quadratic in $\bi{\hat n}$ (i.e., it has a
  quadrupole). {\it Velocity gradients in the photon-baryon fluid
lead to a quadrupole component of the intensity distribution, which,
through Thomson scattering, is converted into polarization}.  

The polarization of the scattered radiation field, expressed in terms
of the Stokes parameters $Q$ and $U$, 
is given by $(Q+iU) \propto \sigma _T\int d\Omega
^{\prime} ({\bf m}\cdot \hat {{\bf n}}^{\prime })^2 T(\hat {{\bf n}%
}^{\prime })$ $ \propto \lambda _p{\bf m}^i{\bf m}^j\partial
_iv_j|_{\tau_R}$, 
where $\sigma _T$ is the Thomson scattering cross-section and we have
written the
scattering matrix as $P({\bf m},\hat {{\bf n}}^{\prime })=-3/4\sigma _T({\bf %
m}\cdot \hat {{\bf n}}^{\prime })^2$, with ${\bf m}=\hat {{\bf e}}_1+i\hat {%
{\bf e}}_2$ . In the last step, we integrated over all directions of the
incident photons $\hat {{\bf n}}^{\prime }$.
As photons decouple from the baryons, their mean free path grows very
rapidly, so a more careful analysis is needed to obtain the final
polarization\footnote{The velocity in this
equation is in the conformal gauge.}(Zaldarriaga \& Harari 1995): 
\begin{eqnarray}
(Q+iU)(\hat {{\bf n}})\approx \epsilon \delta \tau _{R}{\bf m}^i
{ \bf m}^j\partial
_iv_j|_{\tau _{R}}, 
\label{polapprox}
\end{eqnarray}
where $\delta \tau _{R}$ is the width of the last-scattering surface and is
giving a measure of the distance photons travel between their last two
scatterings, and $\epsilon$ is a numerical constant that depends on the
shape of the visibility function. 
The appearance of ${\bf m}^i {\bf m}^j$ in Equation \ref{polapprox} assures
that $(Q+iU)$ transforms correctly under rotations of $(\hat {{\bf e}%
}_1,\hat {{\bf e}}_2)$. 

If we evaluate Equation 
\ref{polapprox} for each Fourier mode and combine them
to obtain the total power, we get the equivalent of Equation
\ref{approxcl},
\begin{eqnarray}\label{approxcel}
  l(l+1)C_{El}\approx A \epsilon^2 (1+3R)^2 (k^*\delta\tau_R)^2  
\sin^2(k^*c_s\tau_R),
\end{eqnarray}
where we are assuming $n=1$ and that $l$ is large enough that factors
like $(l+2)! /(l-2)!\approx l^4$. The extra $k^*$ in Equation
\ref{approxcel} originates in the gradient in Equation
\ref{polapprox}. As will be discussed later, density perturbations
produce no $B$, so only three power spectra are needed to characterize
the maps.

The curves in Figure  \ref{figclcdm} illustrate the differences between
temperature and polarization power spectra. The large-angular scale
polarization is greatly suppressed by the $k\delta\tau_R$ factor. 
Correlations over large angles can only be created by the long-wavelength 
perturbations, but these cannot produce a large
polarization signal because of the tight coupling between photons and
electrons prior to recombination. Multiple scatterings make the
plasma very homogeneous; only wavelengths that are small enough to
produce anisotropies over the  mean free path of the photons will give
rise to a significant quadrupole in the temperature distribution, and
thus to polarization. Wavelengths much smaller than the mean free path
decay due to photon diffusion (Silk damping) and so are unable to
create a large quadrupole and polarization. As a result polarization
peaks at the scale of the mean free path. 

On sub-degree  angular 
scales, temperature, polarization, and the cross-correlation power
spectra show
acoustic oscillations (Fig. \ref{figclcdm}).
In the polarization and cross-correlation spectra the peaks are much sharper. 
The polarization is produced
by  velocity gradients of the photon-baryon fluid
at the last-scattering surface (Equ. \ref{polapprox}). 
The temperature receives
contributions from density and velocity perturbations (Equ.
\ref{approxcl}), and  the oscillations in each partially
cancel one another, making the features in the temperature spectrum less
sharp. The dominant contribution to the temperature comes from the 
oscillations in the density (Equ. \ref{solution1}), 
which are out of phase with the velocity.
This explains the difference in location between the temperature and
polarization peaks. The extra gradient in the polarization signal,
Equation 
\ref{polapprox}, explains why its overall amplitude peaks at a smaller
angular scale.

\subsection{The Information Encoded in the Polarization Generated at the LSS}

Now that we have reviewed the physics at recombination, we consider
what can be learned from a measurement of polarization on
sub-degree angular scales. 
The temperature and polarization anisotropies produced by density
perturbations are characterized by three power spectra.
Once polarization is measured, we have
two extra sources of information from which to extract constraints on
parameters. Although constraints should get tighter in principle, in
practice 
because polarization anisotropies are significantly smaller than
temperature ones, and thus more difficult to detect, 
parameter constraints do not improve that much from the measurement of
the polarization. This is
specially true if we restrict ourselves to the standard cosmological
parameters such as those describing the matter content of the Universe, 
but not necessarily true for others or when some degeneracies are
considered (Zaldarriaga et al. 1997; Eisenstein et al. 1999). 
This fact can be viewed in two different ways. One could
say that a measurement of polarization in the acoustic peak region 
is not as relevant because it would not dramatically alter our
constraints on parameters.  Alternatively, one could say that it will provide 
a consistency check because, once the 
parameters of the model are determined from the temperature
anisotropies, polarization is accurately predicted. We should keep in
mind that, although everything seems to be falling in line with
theoretical predictions, there are remarkably few real consistency
checks, when the same quantity is measured accurately in two
independent ways. 
We will  mention three consistency checks that can be performed using
polarization on sub-degree angular scales. 

In what is quickly becoming the standard model, perturbations are
generated during a period of inflation.  The potential of CMB
anisotropies to test inflationary models has long been recognized
(e.g., Hu \& White 1996; Hu, Spergel, \& White 1997). One of the main
predictions of inflation is that perturbations are correlated over
scales larger than the horizon at recombination and that modes have
phase coherence, which leads to the peaks in the power spectrum.
Because the speed of the acoustic waves is smaller than the speed of
light, one can produce the acoustic peaks in the temperature and
polarization spectra without resorting to above-horizon
perturbations (e.g., Turok 1996a, b).
Moreover, because temperature can be produced after recombination (when
the horizon is bigger) through the ISW effect, correlations on scales
larger than two degrees (the scale of the recombination horizon) 
are not proof of above-horizon perturbations.
Correlations of the polarization, on the other hand, provide a 
clearer test because there is no equivalent to
the ISW effect (Spergel \& Zaldarriaga 1997). This test has now been performed 
by the {\it WMAP} team, further strengthening the case for inflation 
(Pieris et al. 2003)

It is customary to assume that the power spectrum of primordial
fluctuations is a simple power law, perhaps with a logarithmically
varying spectral index. Although this is well motivated in the context
of inflation, it should be checked.  Moreover, preliminary determinations of 
the power spectrum, combining results from {\it WMAP} and large-scale 
structure, hint that something strange may be going on.
Constraints on the power spectrum
of primordial fluctuations come from comparing the amplitude of
temperature perturbations at different angular scales. At least some
of the changes produced by differences in the primordial spectrum 
can be mimicked by changes in the cosmological
parameters. If both temperature and polarization are measured, there
are three independent measures of the level of fluctuations 
on each scale. Moreover, temperature and
polarization depend differently on cosmological parameters. Thus, the
simultaneous measurement of temperature and polarization can be used
to separate the early (inflation era) 
and later Universe physics affecting the CMB
(e.g., Tegmark \& Zaldarriaga 2002).

In the simplest inflationary models, perturbations only come in the
adiabatic form. However, in more complicated inflationary models or in
other classes of early-universe ``scenarios,'' isocurvature
perturbations can arise. In a model with photons, baryons,
neutrinos and cold dark matter, there are actually four isocurvature modes
in addition to the adiabatic one (e.g., Bucher, Moodley, \& Turok 2000; 
Rebhan \& Schwarz 1994; Challinor \& Lasenby 1999). The
simultaneous measurement of temperature and polarization will allow us 
to put constraints on small admixtures of these components
 that would be impossible otherwise (Bucher, Moodley, \& Turok 2001). 
 
 Finally, because polarization perturbations are accurately predicted
 once temperature is measured, polarization can be used to make
 several consistency checks on the way recombination happened. In
 particular, changes in fundamental constants such as the fine
 structure constant, the gravitational constant, the speed at which
 the Universe was expanding during recombination, or the presence of
 ionizing photons on top of those from the CMB should be severely
 constrained once the data are in (Kaplinghat, Scherrer, \& Turner 1999; 
Peebles, Seager, \& Hu 2000; Hannestad \& Scherrer 2001; Landau, Harari, 
\& Zaldarriaga 2001; Uzan 2003; Zahn \& Zaldarriaga 2003).

\section{The $E-B$ Decomposition and the Imprint of Gravity Waves}\label{gw}

The study of the anisotropies produced by GW has a
long history (e.g., Polnarev 1985; Crittenden, Davis, \& Steinhardt 1993; 
Coulson, Crittenden, \& Turok 1994; Crittenden, Coulson, \& Turok 1995). 
Perhaps the biggest driving force behind the push to measure
polarization is the possibility of detecting a stochastic background
of GW. The same mechanism that creates the seeds for
structure formation, the stretching of vacuum fluctuations during inflation, 
is also
expected to generate a stochastic background of GW.
The amplitude of the GW in the 
background is directly proportional to the Hubble constant during inflation, or equivalently the square of energy scale of inflation. 

As photons travel in the metric perturbed by a GW 
[$ds^2=a^2(\tau)$ $[-d\tau^2 $ $ +(\delta_{ij}+ h^{T}_{ij})dx^i dx^j]$],
they get redshifted or blueshifted depending on their direction of
propagation relative to the direction of propagation of the GW
and the polarization of the GW. For example, for a
GW traveling along the $z$ axis, the frequency shift is
given by ${1 \over \nu}{d \nu \over d\tau}= {1 \over 2}\ \hat
{n}^i\hat {n}^j {\dot h}^{T(\pm)}_{ij}= {1 \over 2}\ (1-\cos^2
\theta)e^{\pm i 2\phi}\ \ \dot h_t\ \exp(i\bi k\cdot \bi x)$, where
$(\theta,\phi)$ describe the direction of propagation of the photon,
the $\pm$ correspond to the different polarizations of the GW, 
and $h_t$ gives the time-dependent amplitude of the GW.
During the matter-dominated era, for example, 
$h_t= 3 j_1(k\tau)/k\tau$.
This effect is analogous to the ISW effect: time changes in the metric
lead to frequency shifts (or equivalently shifts in the temperature of
the black body spectrum). Notice that the angular dependence of this
frequency shift is quadrupolar in nature. As a result, the temperature
fluctuations induced by this effect as photons travel between
successive scatterings before recombination 
produce a quadrupole intensity distribution, which,
through Thomson scattering, lead to polarization. {\it We want to stress
that polarization is always generated by Thomson scatterings, whether
we talk about polarization generated by density perturbations or by
GW. All that differs in these two cases is what is
responsible for the quadrupole anisotropies.}

Figure \ref{figclcdm} shows the temperature and polarization anisotropy
power spectrum produced by GW. We have arbitrarily normalized 
the GW contribution so that the ratio of $l=2$ contributions
from tensors and density perturbations $T/S \approx 0.1$, corresponding
to an energy scale of inflation, $E_{\rm inf}\approx  2\times 10^{16}\, {\rm
GeV}$. The temperature power spectrum, produced mainly after
recombination, is roughly flat up to $l\approx 100$, and then rapidly falls
off. By contrast, polarization is produced at recombination (except for the
reionization bump at low values of $l$, to be discussed later). 
The power spectrum of polarization peaks at $l\approx  100$ because, 
just as for density perturbation, GW much larger than the
mean free path of the photons at decoupling cannot generate an
appreciable quadrupole. The spectrum falls on small scales as well
because the amplitude of GW decays when
they enter the horizon, as they redshift away. 
The decay occurs at horizon crossing, much before the Silk
damping scale.

So far we have ignored the spin 2 nature of polarization. Both $E$ and $B$ 
power spectrum are generated by GW (Kamionkowski et al. 1997a; Seljak \& 
Zaldarriaga 1997). The current push to improve polarization measurements
follows from the fact that density perturbations, to linear order in
perturbation theory, cannot create any $B$-type polarization. 
As a rough rule of thumb, the amplitude of the peak in the $B$-mode
power spectrum for GW is
$[{l(l+1)C_{Bl}/ 2\pi }]^{1/2}=0.024 ({E_{\rm inf}/ 10^{16}{\rm
GeV}})^2 \mu \rm K$.

One can understand why GW can produce $B$, 
while density perturbations cannot, by analyzing the
symmetries of the problem. We recall that we are dealing with a linear
problem, so one Fourier mode of density perturbations or a single
GW needs to be analyzed. When calculating the temperature
and polarization maps that would be observed in a universe with a 
single Fourier mode of the density, one realizes that the problem
has two symmetries. There is symmetry of rotation around the $\bi k$ 
vector and symmetry of reflection along any plane that contains the
$\bi k$ vector. The pattern of polarization produced by this mode has
to respect these two symmetries, which in turn implies that there
cannot be any $B$ modes generated in this case. In Figure \ref{eb1}, if
the cross represents the tip of the $\bi k$ vector, it is clear that
only the $E$ patterns respect the symmetries. For one polarization
of the GW, these symmetries are not satisfied, as is obvious
from the pattern of redshifts they produce, ${1 \over \nu}{d \nu \over
d\tau} \propto {1 \over 2}\ (1-\cos^2
\theta)e^{\pm i 2\phi}$. Gravitational
waves can produce a $B$ pattern.

\section{After Recombination}\label{afterrec}

In this section we will discuss a few different effects that can change the
polarization pattern after recombination.

\subsection{Reionization}

So far we have ignored the fact that hydrogen became ionized when the
first sources of UV radiation started shining. The exact time and way
this happened is not yet fully understood, the recent results from {\it WMAP}
indicate that reionization started pretty early, with a prefer value of 
$z\approx 20$ (Bennett et al. 2003). There are also
indications from studies of absorption toward high-redshift quasars 
that something happened around redshift $z\approx 6-8$ (Fan et
al. 2002).  It could well be that reionization
is not a one-stage process. Depending on the details of
the sources of radiation,  a more complicated ionization history could have 
occurred (e.g., Wyithe \& Loeb 2003). 

In any event, at these redshifts the density of hydrogen was so much
lower than at recombination that even though hydrogen is fully
ionized the optical depth is still relatively low; somewhere around 
$17\%$ is what {\it WMAP} favors. Even though only a fraction of the photons 
scatter after recombination, this still has a dramatic effect on polarization
on large angular scales, of order 10 degrees. On these scales the
polarization produced at recombination is minimal because these scales
are much larger than the mean free path at recombination. As can be
seen in Figure \ref{figclcdm}, polarization has a power-law decay toward
large scales. 

After recombination, photons were able to travel a significant distance
over which the large-wavelength modes can induce a quadrupole. This
quadrupole leads to a bump in the polarization power spectrum shown in
Figure \ref{figclcdm} that is
orders of magnitude higher than what polarization would be if it 
were not for reionization (e.g., Zaldarriaga 1997). If there were no
reionization the power law decay would continue up to $l=2$.  

A detection of this large-scale signal will allow a precise
measurement of the total optical depth and provide some additional
information on how and when reionization happened. The height of the
reionization bump is proportional to the total optical depth, while the
shape and location of the peak contains information about when reionization
happened (Zaldarriaga 1997; Kaplinghat et al. 2003a). 
The total amount of information that can be extracted is,
however, limited because at these large scales there are only a few
multipoles that one can measure (i.e., cosmic variance is large; e.g., Hu 
\& Holder 2003). However, a measurement of the optical depth is very important 
because it breaks degeneracies between determinations of several parameters
(Zaldarriaga et al. 1997; Eisenstein et al. 1999). This excess of large-scale 
polarization is what {\it WMAP} used to determine a redshift of reionization 
around $z\approx 20$ (Spergel et al. 2003). 

\subsection{Weak Lensing}

As photons propagate from the last-scattering surface they get
gravitationally deflected by mass concentrations, the large-scale
structure of the Universe. This gravitational
lensing effect changes both temperature and polarization anisotropies
(e.g., Seljak 1996; Bernardeau 1997, 1998; Hu 2000), 
but has rather profound consequences for the pattern of CMB
polarization. Even a polarization pattern that did not have any $B$
component at recombination will acquire some $B$ due to gravitational
lensing (Zaldarriaga \& Seljak 1998). 
The effect is simple to understand. Because of lensing, a
photon originally traveling in direction $\bi {\hat n}$ will be observed
toward direction $\bi{\hat  n}^{\prime}$. We can use the analogy with a vector
field and assume we start with the pattern that is a perfect
gradient. As vectors are slightly shifted around due to
lensing, the field will develop a curl component. 

In Figure \ref{figclcdm} we show the $B$ component generated by lensing of
the $E$ mode. It is clear from the figure that if the level of the
GW background is much lower than shown there, 
then the lensing signal would be larger at almost all scales. 
Note that the lensing $B$ does not have
the reionization bump at low $l$. This is so because the power is
actually coming from ``aliasing'' of the small-scale polarization
power rather than from a rearrangement of the original large-scale
power.

The ultimate limitation for detecting the stochastic
background of GW comes form the spurious $B$ modes
generated by lensing. The lensing distortions to the temperature and
polarization maps make them non-Gaussian (e.g., Bernardeau 1997, 1998;
Zaldarriaga 2000; Cooray \& Hu 2001; Okamoto \& Hu 2002). 
Methods have been developed to use this non-Gaussianity to measure the
projected mass density (Seljak \& Zaldarriaga 1999a; 
Zaldarriaga \& Seljak 1999; Hu 2001a,b; Okamoto \& Hu 2002),  
which can be used to clean, at least partially, 
this contamination. With the methods proposed so far,
the lowest energy scale that seems measurable is $E_{\inf}=2\times 10^{15}\,
{\rm GeV}$ (Kesden, Cooray, \& Kamionkowski 2002; Knox \& Song 2002). 
The energy scale of inflation is at present only very loosely
constrained, so it is perfectly possible that the GW background is too
small to be observed. However, if inflation is related to the Grand 
Unification scale around $10^{16}\ {\rm GeV}$, then there is a good chance 
that the GW background will be detected. 

The lensing effect is not only a nuisance for detecting GW;
it is interesting to constraint the large-scale structure that
is doing the lensing. Because the last-scattering surface is at such
high redshift, lensing of the CMB may eventually provide one of the
deepest probes for large-scale structure. 
The level of structure on scales of order $2-1000 $
$\rm Mpc$ at redshifts from $z\approx 10$ to $z\approx 0$ may eventually be
constrained with this technique (e.g., Zaldarriaga \& Seljak 1999). 
Lensing will not only allow a measurement of the 
level of fluctuations but may lead to actual
reconstructed maps of the projected mass density 
that can be correlated with maps produced by CMB experiments and 
other probes (e.g., Goldeberg \& Spergel 1999; Seljak \& Zaldarriaga 1999b; 
Spergel \& Goldberg 1999; Van Waerbeke, Bernardeau, \& Benabed 2000; Benabed, 
Bernardeau, \& van Waerbeke 2001; Song et al. 2003) and to strong constraints 
on parameters such as the mass of the neutrinos (Kaplinghat, Knox, \& Song 
2003b). On small scales, large mass concentrations such as cluster
may leave a detectable signature (Seljak \& Zaldarriaga 2000). 

\subsection{Foregrounds}

As we have discussed, a significant amount of information is encoded
in the large-scale polarization. The ability of polarization to
determine the reionization history and with that help break some of
the degeneracies between cosmological parameters relies on the
large scales (Zaldarriaga et al.  1997).

Measuring polarization over the whole sky might be a problem
when it comes to foregrounds. First, it means that one may have to use
patches of the sky that are not that foreground free. Second, it
appears that at least for the unpolarized component galactic 
foregrounds have rather red spectra, affecting large scales the
most.  

At microwave frequencies synchrotron, free-free, and dust emission are 
foreground contaminants. At present we have a fairly good understanding of
the unpolarized component of the emission from our Milky Way 
and from distant galaxies (usually referred to as point sources). 

The situation with regard to polarization is far worse. Both
synchrotron and dust emission are expected to be polarized. In the
case of synchrotron the theoretical maximum is 70\%. Most of our
knowledge about synchrotron polarization comes from surveys at relatively
low frequencies. Extrapolation to frequencies where CMB
experiments operate  is
severely hampered by Faraday rotation. Moreover, most modern surveys
have concentrated on regions near the Galactic plane. How to
extrapolate to higher latitudes remains unclear. 

We will not dwell on foregrounds much longer because at this point
there is not that much we can say.  Time will tell if we are lucky
again, as with the temperature anisotropies, or if foregrounds will
spoil the potential fun of studying polarization.  The interested
reader is referred to the articles in the volume edited by  
de~Oliveira-Costa \& Tegmark (1999). More recent analysis of existing
maps, some of which deal with the issue of whether foregrounds
contaminate equally the $E$ and $B$ components can be found in a number of 
recent studies
(Tucci et al. 2000, 2002; Baccigalupi et al. 2001; Bruscoli et al. 2002; 
Burigana \& La Porta 2002; Giardino et al. 2002).
A recent summary including constraints coming from the PIQUE
experiment is presented in 
 de~Oliveira-Costa et al. (2003b). In the case of the 
DASI detection, the spectral index of the temperature anisotropies 
was very tightly constrained (ruling out any significant contamination there),
and the temperature and polarization maps
were correlated. This, together with the (rather  weak) spectral index
constraints from the polarization data, argued against any significant
foreground contamination (Kovac et al. 2002).

\subsection{$E-B$ Mixing: Systematic Effects}

There are a variety of systematic effects that lead to mixing between
$E$ and $B$ modes. In finite patches of sky, the separation cannot be
done perfectly (Lewis, Challinor, \& Turok 2002; Bunn et al. 2003). 
This is analogous to trying to decompose a vector
field into its gradient and curl parts when it is measured on a finite
part of a plane. Vector fields that are gradients of a scalar with zero
Laplacian will have both zero curl and divergence. 
In fact, in a finite patch one can construct
a basis for the polarization field in which basis vectors can be split
into three categories. There are pure $E$, pure $B$, and a third
category of modes that is ambiguous, which receives contributions from
both $E$ and $B$. The number of ambiguous modes is proportional the
number of pixels along the boundary. 

Aliasing due to pixelization mixes $E$ and $B$; the power that is
aliased from sub-pixel scales leaks into both $E$ and $B$. This is
particularly important because $E$ polarization is expected to be much
larger than $B$ modes, and because the $E$ polarization power spectra
is relatively very blue.

Finally, other effects such as common mode and differential
gain fluctuations, line cross-coupling, pointing errors, and
differential polarized beam effects will  create a spurious $B$
signal from temperature and/or $E$ components (Hu, Hedman, \& Zaldarriaga 2003).

\section{Conclusions}\label{conc}

We have summarized the physics behind the generation of a small degree
of polarization in the CMB. Quadrupole
anisotropies in the radiation intensity 
at the last-scattering surface through Thomson scattering
lead to a small degree of linear polarization. These quadrupole
anisotropies can be generated by both density perturbations (mainly
through ``free streaming" of the Doppler effect) and by GW.

The quadrupole generated by GW leads to a distinct
pattern of polarization on the sky. Such a pattern has a curl
component, and thus the CMB polarization can serve as an indirect
GW detector. If inflation is the source of the density
perturbations, it is also expected to generate a stochastic background
of GW. Searching for this background through CMB
polarization has become one of the driving forces for the field. The
level of the $B$ component produced by GW is expected to be quite
small, so the first generation of polarization experiments should see
$B$ modes consistent with zero.  

After recombination several processes can affect
polarization. Gravitational lensing distorts the polarization patter
generating a $B$ component, even in the absence of GW.
This source of noise could become the final limit to the detectability 
of the GW background. 

The reionization of the Universe provides a new opportunity for the
CMB photons to scatter. It leaves a signature in the large-scale
polarization, a bump in the power spectrum. If detected, it would help
constrain the epoch of reionization and break many of the 
degeneracies that occur in CMB fits of cosmological parameters.

The DASI experiment has recently detected polarization. It found a
pattern of polarization consistent with having no $B$ modes, just as
expected. It also favors a spectrum that rises toward small scales, 
just as the theory predicts. Moreover, it provides a tentative
detection of a cross correlation between temperature and polarization
at the level predicted by the model. The {\it WMAP} satellite already released 
a high signal-to-noise ratio measurement of the cross correlation between 
temperature and polarization. These results  lead to the conclusion that the 
Universe reionized at a surprisingly high redshift and provided further 
evidence in favor of inflation.  Everything is looking good.  Time will tell 
if more sensitive polarization experiments will eventually
fulfill their promise and help us solve some of the remaining
mysteries about our Universe.

\end{document}